\input{aipcheck}

\documentclass[
    ,final            % use final for the camera ready runs
  ]
  {aipproc}

\layoutstyle{8x11double}

\def\lesssim{\mathrel{\hbox{\rlap{\hbox{\lower4pt\hbox{$\sim$}}}\hbox{$<$}}}}
\def\gtrsim{\mathrel{\hbox{\rlap{\hbox{\lower4pt\hbox{$\sim$}}}\hbox{$>$}}}}

\begin{document}

\title{The Inhomogeneous Background of $\rm H_2$ Dissociating
  Radiation During Cosmic Reionization} 



\author{Kyungjin Ahn}{address={Department of Earth Science, Chosun
  University, Gwangju 501-759, Korea}
,altaddress={Department of Astronomy, 1 University Station, C1400,
  Austin TX 78712}
} 
\author{Paul R. Shapiro}{address={Department of Astronomy, 1 University Station, C1400,
  Austin TX 78712}
}
\author{Ilian T. Iliev}{address={Universitaet Zuerich,
Institut fuer Theoretische Physik,
Winterthurerstrasse 190
CH-8057 Zuerich, Switzerland}
}  
\author{Garrelt Mellema}{address={
Department of Astronomy, Stockholm University
SE-106 91 Stockholm, Sweden}
}
\author{Ue-Li Pen}{address={Canadian Institute for Theoretical
  Astrophysics, University of Toronto, 60 St. George St., Toronto, ON
  M5S 3H8, Canada}
}

\begin{abstract}
The first, self-consistent calculations of the cosmological $\rm H_2$
dissociating UV background produced during the epoch of 
reionization (EOR) by the sources of reionization are
presented. Large-scale radiative transfer simulations of  
reionization trace the impact of all the ionizing starlight on the IGM
from all the sources in our simulation volume down  
to dwarf galaxies of mass $\sim 10^8 M_\odot$, identified by very
high-resolution N-body simulations, including the self-regulating  
effect of IGM photoheating on dwarf galaxy formation. The UV continuum
emitted below 13.6 eV by each source is then  
transferred through the same IGM, attenuated by atomic H Lyman series
resonance lines, to predict the evolution of the  
inhomogeneous background in the Lyman-Werner band of $\rm H_2$ between
11 and 13.6 eV.  
\end{abstract}

\maketitle

\section{Suppression of Minihalo Pop III Star Formation by $\rm H_2$ Dissociating UV 
Background}
\label{sec:suppression_MH} 
Simulations suggest that first stars formed inside minihalos of mass $M\sim
10^6 M_\odot$ at $z\gtrsim 20$, when  
$\rm H_2$ cooled the primordial, metal-free halo gas and gravitational
collapse ensued. Formation of first stars inside minihalos, however,
is suppressed when $\rm H_2$ cooling is suppressed due to a strong
$\rm H_2$ Lyman-Werner(LW) band radiation field. This process occurs
when the LW intensity $J_{\rm LW}$ exceeds the threshold LW intensity
$(J_{\rm LW})_{\rm threshold}$, which is usually set by requiring that
cooling time equals the dynamical time  (e.g. Haiman, Rees \& Loeb 1997). 

It is important, therefore, to correctly calculate the rise of $\rm
H_2$ dissociating background in the Universe.
Earlier estimates (e.g. Haiman, Abel \& Rees 2000) found that sources of
reionization made  
$J_{\rm LW} >(J_{\rm LW})_{\rm threshold}$ long before
reionization is complete, thus sterilizing  minihalos before they could  
contribute significant reionization.

The LW background intensity would show spatial fluctuation due to
the inhomogeneous distribution of reionization sources. However,
previous calculations 
are based upon a homogeneous universe approximation. In these calculations,
the sources and the IGM opacity were both uniformly distributed, in
space with uniform emissivity given either by analytical approximation
(e.g. Haiman, Abel \& Rees 2000), or by sum over sources found in small-box simulations,
too small to account for large-scale clustering of sources or follow
global reionization (e.g. Ricotti, Gnedin \& Shull 2002; Yoshida et
al. 2003).

Here we present the first self-consistent radiative transfer
calculations of the  
inhomogeneous LW background produced by the same sources which reionized the 
universe in a large-scale radiative transfer simulation of
reionization. We show that the fluctuation is significant enough to
induce widely-different evolutionary history in the first star
formation in different parts of the Universe.

%\section{The First Self-Consistent Calculation of the Inhomogeneous LW
%  Dissociating Background during Epoch of Reionization (EOR) }
\section{A Method to Calculate the
Inhomogeneous LW Background during Epoch of Reionization}
\label{sec:picketfence}

\begin{figure}[ht]
\includegraphics[height=.22\textheight]{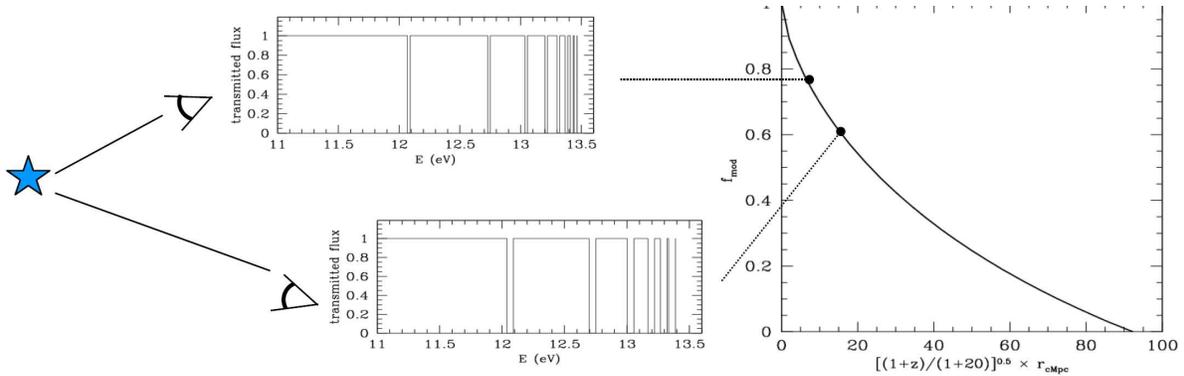}
\caption{``picket-fence'' modulation. A source observed at different
places suffers different amounts of attenuation, dilution, and
redshift. As continuum (flat-spectrum assumed) photons
redshift, all the photons which  
have redshifted to H atom resonance lines are turned into low-energy
photons. This leaves gaps  
in observed spectrum, which depend on the distance to the source. The
modulation factor (right panel)  
is the effective transmission rate, which drops as the distance
increases, until source is completely  
attenuated at $r_{\rm cMpc} = 92 \,[(1+z)/21]^{-0.5}$. With this
modulation factor, we are able to  
compute LW intensity without costly multi-frequency calculation. }
\label{fig-picketfence}
\end{figure}

Calculating the inhomogeneous build-up of the LW background is
computationally challenging. First,
the mean free path for LW photons ($\sim$ 100 cMpc) is much larger
than the mean free path for H  
ionizing photons. We need to account for sources distributed over large
volume and look-back time. Second,
LW band photons redshift and get attenuated by H atom Lyman series
resonance lines as they travel through the expanding universe. This requires 
a multi-frequency radiative transfer calculation in
a cosmological volume ($ \gtrsim (100 \,\rm cMpc)^3$), which is 
very expensive computationally.

%Attenuation of LW band photons occurs in the following way.
%As a LW band photon  
%travels, when its frequency redshifts into an H resonance line, it is
%absorbed and some fraction turn into  
%low frequency photons. If it resonantly scatters, it is quickly
%reabsorbed, until all resonant photons turn  
%into low frequency photons below LW bands. For homogeneous universe,
%this gives ``saw-tooth'' 
%modulation of the spectrum (HRL, HAR). But in inhomogeneous universe, this
%treatment is not valid. 

We have developed a simple, yet accurate method to calculate the
radiative transfer of LW band photons, which alleviates the need for a
multi-frequency calculation. This is achieved by using a pre-computed
modulation 
factor, expressed in terms of the comoving distance between a source
and an observer, that accounts for the attenuation of LW band photons from a
single source. We call this a ``picket-fence'' modulation factor
$f_{\rm mod}$, after the shape of 
the LW intensity $J_\nu$ which is composed of both attenuated and
unattenuated parts (Fig. 1). The picket-fence modulation factor is
computed by averaging a normalized $J_\nu$ (at $r_{\rm cMpc}(z_{\rm
s},\,z_{\rm obs})$) 
over frequency in the range of [11.5 -- 13.6] eV. 

At a given point in space, one then looks back in time to search for
sources located inside the LW band horizon ($\sim$ 100 cMpc). For each
source found this way, a ``bare'' flux is obtained from its luminosity
$L_\nu$ and 
the luminosity distance,  which corresponds to a flux in an  optically
thin limit. This is then  multiplied by $f_{\rm mod}$ to
account for the attenuation by H atom Lyman series resonance
lines. Since $f_{\rm mod}$ is a function of the comoving
distance only, an expensive multi-frequency calculation is not
required. When looking back in time for contributing sources, we draw
past light-cones in order to account for the retarded time effect due
to finite light-crossing time
(Fig. 2). Sources are distributed in a discrete sense in time,
following  our reionization simulations results (e.g. Iliev et al. 2007).

\begin{figure}[ht]
\includegraphics[height=.25\textheight]{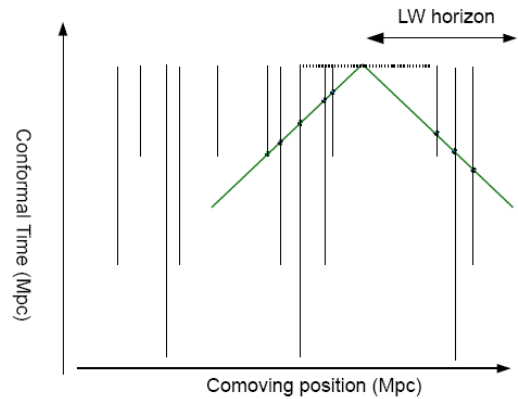}
\caption{Conformal spacetime diagram, to be used to calculate 
source contribution from cosmological distances. Each 
source contributes LW flux of $f_{\rm mod} L/D^{2}_{\rm L}$, 
where L is luminosity, $D_{\rm L}$ is the 
luminosity distance to the source, and $f_{\rm mod}$ is the picketfence 
modulation factor which accounts for IGM opacity. This scheme
naturally accounts  
for the finite light-crossing time of radiation. }
\label{fig-conformal}
\end{figure}

\section{Inhomogeneous LW Background}
In order to calculate inhomogeneous LW background, we use
``self-regulated'' reionization simulation results by 
Iliev et al. (2007). Their radiative transfer calculation is based upon
an N-body 
simulation ($1624^3$ particles and $3248^3$ cells) result which
provides the source catalogue for H-ionizing radiative transfer
calculation. This resolves
halos of mass  $M/M_\odot \gtrsim 10^8$, thus accounting for
all atomic 
cooling halos in the $(50 \, {\rm  cMpc})^3$ volume. 
Source formation is self-regulated:
formation of sources inside small-mass halos ($10^8 < M/M_\odot <
10^9$) is suppressed if H II regions overtake their  
formation sites. 
We then use a
coarse-grained mesh ($203^3$ cells) to calculate the fluctuating LW radiation
field. To account for the large LW band horizon, we distribute sources 
periodically around the box where the LW background is calculated.

\begin{figure}[ht]
\includegraphics[height=.23\textheight]{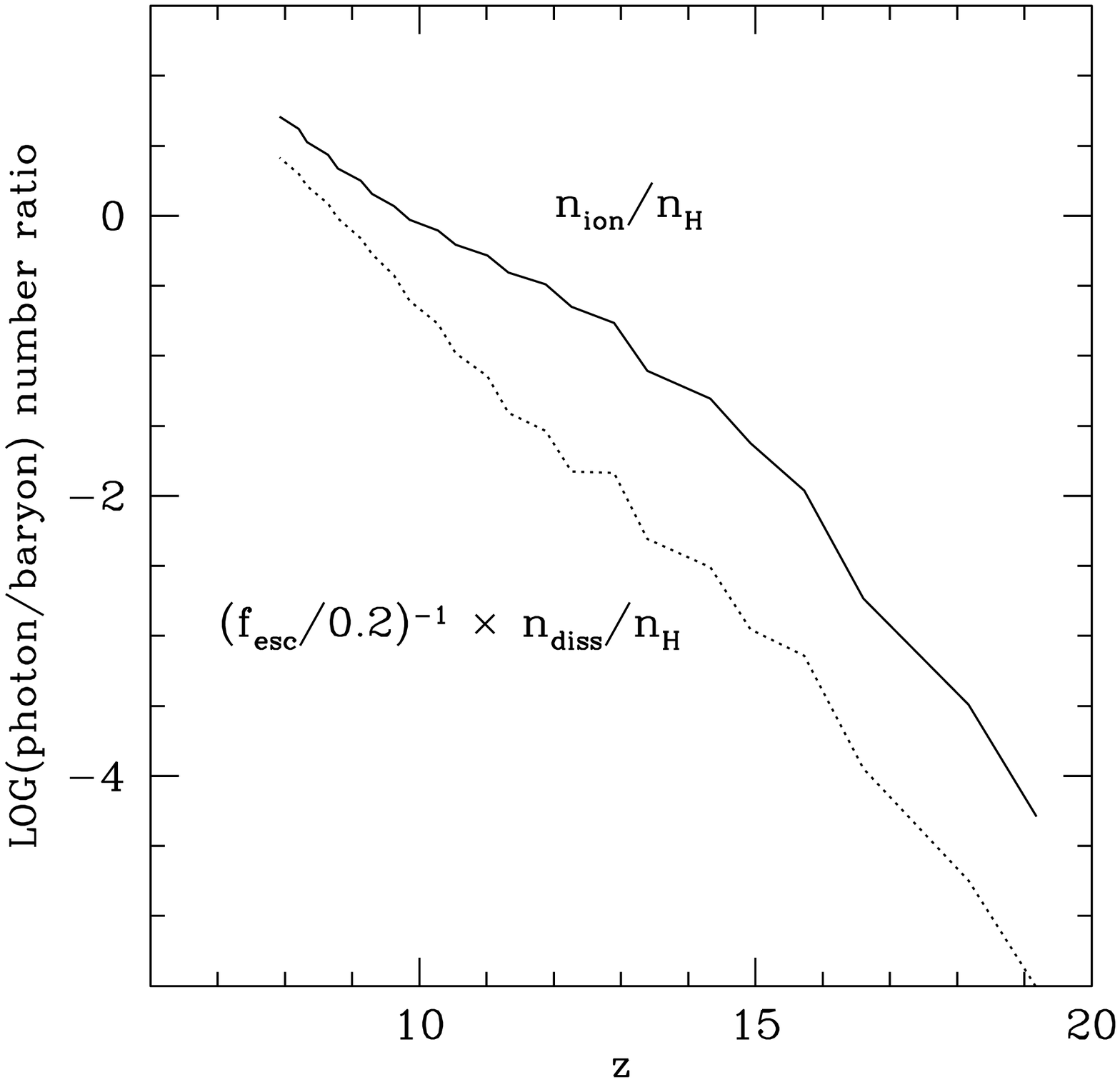}
\includegraphics[height=.23\textheight]{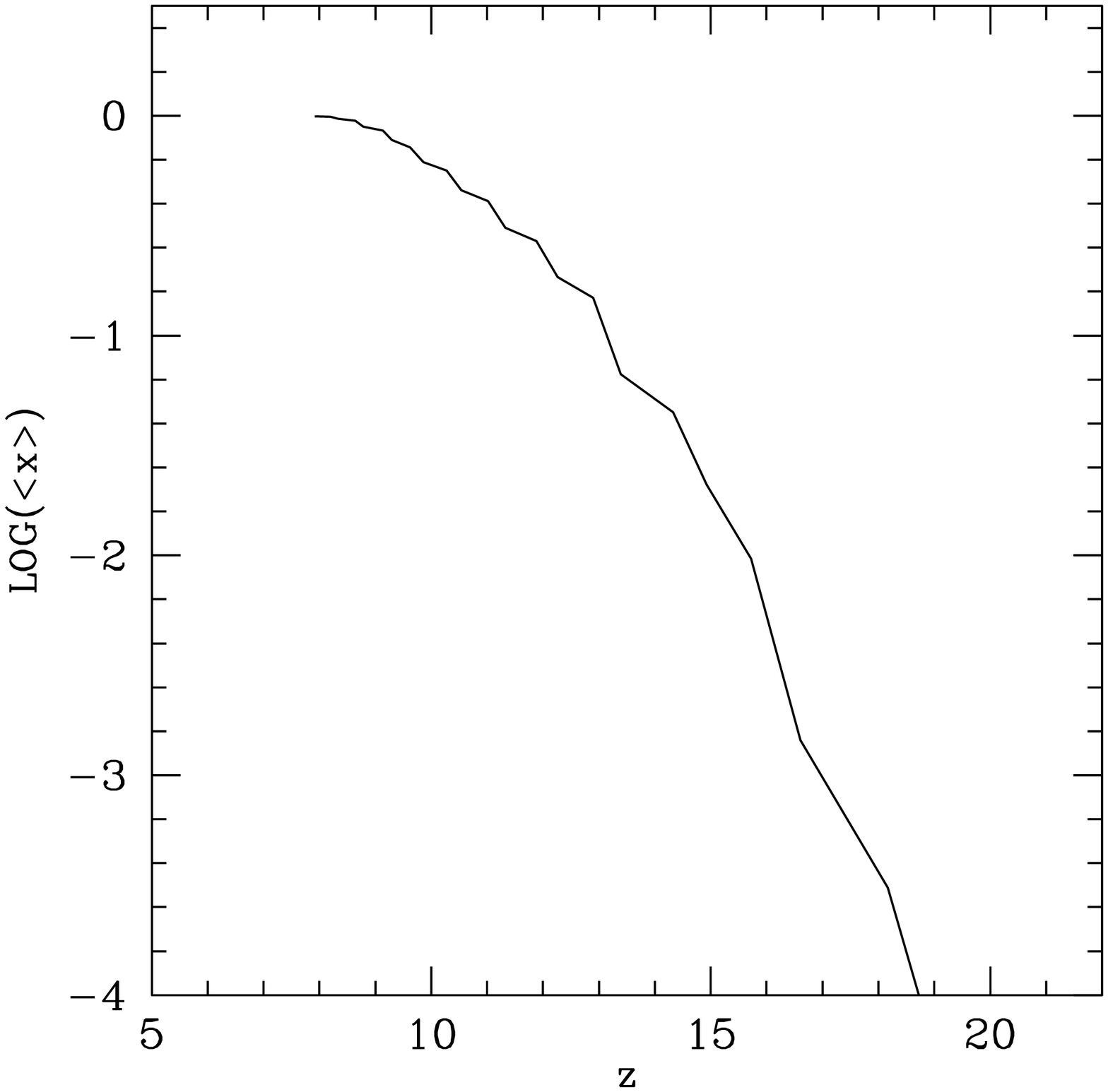}
\includegraphics[height=.23\textheight]{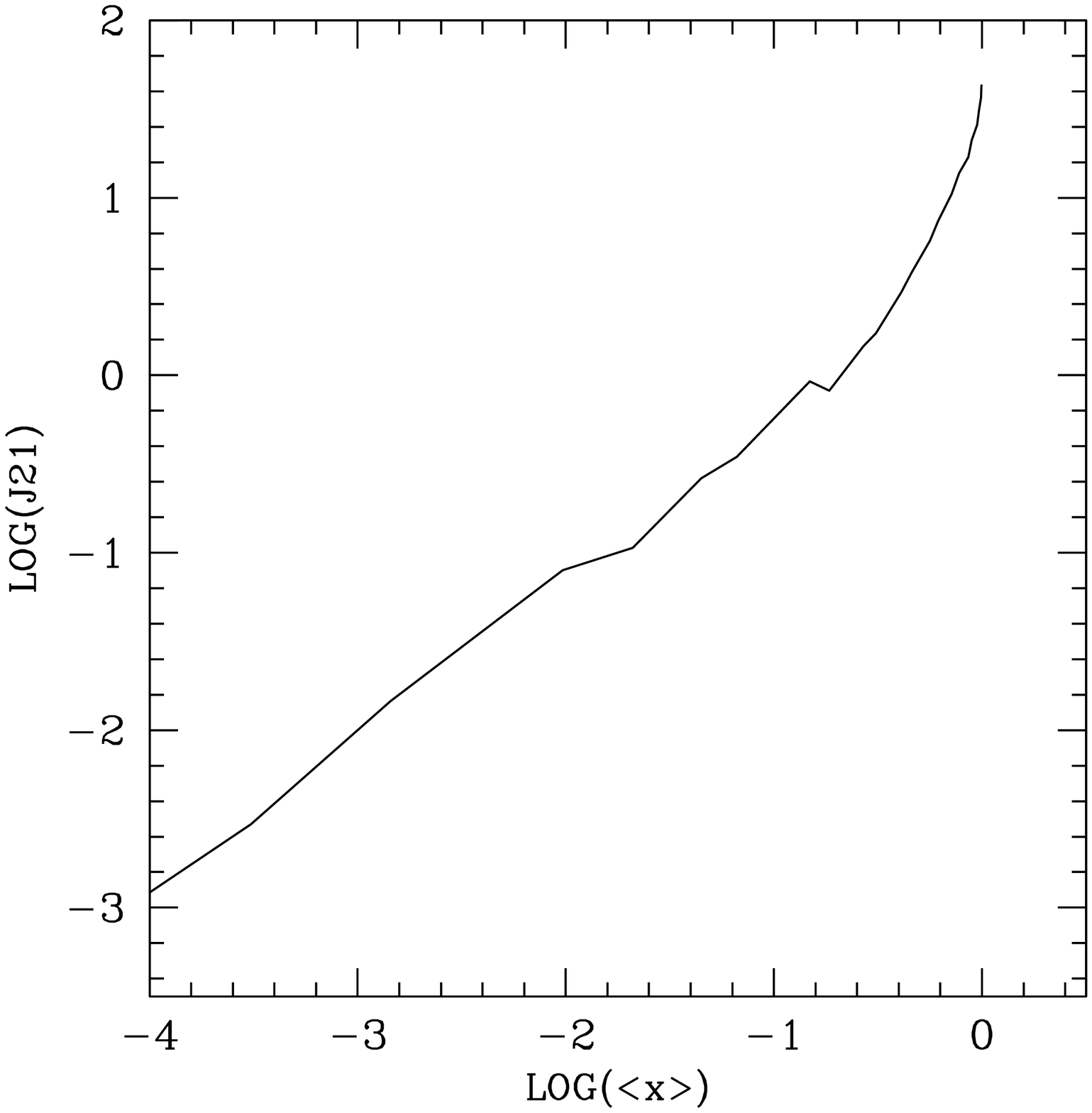}
\caption{Global history of cosmic reionization and LW intensity
evolution. (Left): The accumulated number of ionizing  
photons per baryon vs. the number of dissociating photons per
baron. (Middle): Global evolution of ionization  
fraction $<x>$. (Right): Global evolution of LW intensity
$J_{21}$. Note that if 
reionization source properties change, especially  
in ionizing photon escape fraction ($f_{\rm esc}$) and ratio of
dissociating/ionizing photons, $J_{21}$ may differ even at the same  
global ionized fraction.}
\label{fig-evolution}
\end{figure}

We chose an illustrative case, WMAP3 f2000\_250S, among the reionization
simulation results by Iliev et al. (2007). 
For small mass halos ($10^8 < M/M_\odot <
10^9$), Pop III high efficiency emitters with top-heavy IMF
are assumed, with $f_\gamma = f_{*} \cdot f_{\rm esc} \cdot N_{\rm i}
= 2000$, and with $N_{\rm i}/N_{\rm diss} \simeq 15$, where $f_*$ is
the star formation 
efficiency,  $f_{\rm esc}$ is the ionizing photon escape fraction, $N_{\rm
  i}$ is the number of ionizing  
photons emitted per baryon, and  $N_{\rm diss}$ is the number of $\rm
H_2$ dissociating photons emitted per baryon.
Formation of sources inside large-mass halos ($M > 10^9 M_\odot$ ) are
not suppressed even inside H II regions.  
Pop II low efficiency emitters with Salpeter IMF are assumed, with $f_\gamma
= f_{*} \cdot f_{\rm esc} \cdot N_i = 250$, and with $N_{\rm i}/N_{\rm
  diss} \simeq 1$. 

Fig. 3 summarizes our result on the global ionized
fraction $<x>$ and the mean LW intensity $<J_{21}>$.
We find that $<J_{21}>$ is dominated by small mass
halo contribution initially, while at later epoch by  
large-mass halo contribution in this case. By the time large mass
halos become important, most of small mass halos are already inside H
II regions and get sterilized.  If reionization source properties were
different from our illustrative 
case, however, especially in $f_{\rm esc}$ and $N_{\rm i}/N_{\rm diss}$, the  
same $<x>$ would not mean the same $J_{21}$,
in general. 

A significant inhomogeneity in the LW background is apparent, with
fluctuation scale of a 
few -- $\sim 10 \,{\rm cMpc}$ (Fig. 4). This is due to a highly-clustered
source distribution, which shows a fluctuation scale of a similar kind.
This will induce an inhomogeneous LW feedback,
hence inhomogeneous minihalo star  
formation rate. The most pristine environment will be found where
the most active minihalo star formation activity occurs. Those sites
would exist where the
LW intensity becomes minimum, as shown in Fig. 4. These first-star
forming regions during the epoch of reionization might be detected by
upcoming telescopes such as the James Webb Space Telescope (JWST).

\begin{figure}
\includegraphics[height=.3\textheight]{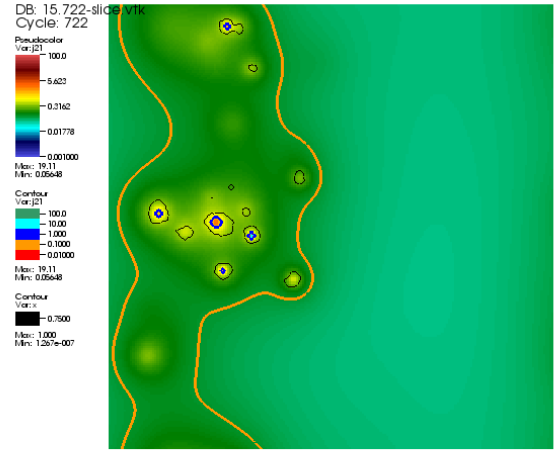}
\caption{Patchy reionization and patchy $\rm H_2$ dissociating
background in $(50 \,{\rm cMpc})^3$ box at $z=15.7$. 
Contours of thick colored lines  
represent different $J_{21}$ contours (orange -- $J_{21}$= 0.1; blue
-- $J_{21}$= 
1), and the black  
contours represent the ionization fronts. 
Minihalos are subject  
to spatially-varying LW feedback effect, thus yielding
spatially-varying minihalo star formation rate. JWST may able  
to detect clustered minihalo population at $z \sim 15$ in regions of
least LW feedback. }
\label{fig-J21color}
\end{figure}

\end{document}